\documentclass[reprint,superscriptaddress,amsmath,amssymb,aps,pra,floatfix,nofootinbib]{revtex4-2}

\usepackage{graphicx}
\usepackage{epstopdf}
\usepackage{float}
\usepackage{physics}
\usepackage{amsmath, amssymb}
\usepackage{amsthm}
\usepackage{dsfont}
\usepackage{tabularx}
\usepackage{units}
\usepackage[dvipsnames]{xcolor}
\usepackage{color,soul}
\usepackage{multirow}

\definecolor{myblue}{named}{NavyBlue}
\definecolor{mygreen}{RGB}{13,255,13}
\usepackage[colorlinks=true,citecolor=myblue,linkcolor=myblue,urlcolor=myblue]{hyperref}
\usepackage[english, capitalise]{cleveref} 

\newcommand{\N}{\mathbb{N}}

\newcommand{\fec}{f_{\text{EC}}}

\theoremstyle{definition}

\theoremstyle{plain}

\theoremstyle{plain}

\theoremstyle{plain}

\theoremstyle{definition}

\begin{document}
	
	\title{Performance of BB84 without decoy states under varying announcement structures}
        
	\affiliation{Institute for Quantum Computing and Department of Physics and Astronomy, University of Waterloo, Waterloo, Ontario N2L 3G1, Canada}

    \author{Zhiyao Wang$^\clubsuit$}
    \email{z2425wan@uwaterloo.ca}
    
    \author{Aodhán Corrigan$^\clubsuit$}
    \email{aodhan.corrigan@uwaterloo.ca}


	\affiliation{Institute for Quantum Computing and Department of Physics and Astronomy, University of Waterloo, Waterloo, Ontario N2L 3G1, Canada}
	\author{Norbert L{\"u}tkenhaus}
  	\email{lutkenhaus.office@uwaterloo.ca}
	\affiliation{Institute for Quantum Computing and Department of Physics and Astronomy, University of Waterloo, Waterloo, Ontario N2L 3G1, Canada}

	\date{\today}
	\begin{abstract}
	{In phase-randomized weak coherent pulse (WCP) implementations of Quantum Key Distribution (QKD) BB84 protocol, the decoy method is often used to compensate BB84's vulnerability against photon number splitting (PNS) attacks. However, this typically introduces extra complexities and requirements on experimental devices. In this paper, we are therefore interested in phase-randomized WCP implementations without the decoy method.} We examine the performance of three QKD protocols with different classical announcement structures, namely BB84, SARG04, and No Public Announcement of Basis (NPAB) BB84, using numerical security proof techniques. We compare secure key rates of the three protocols in asymptotic and finite-size regimes for lossy and noisy channels. The three protocols show different relative advantages depending on the channel behaviour. Canonical BB84 shows robustness against errors and depolarization, SARG04 demonstrates resilience against high loss channels, and NPAB BB84 shows potential advantages against physical misalignment between QKD devices.
	\end{abstract}
	\maketitle
\def\thefootnote{$\clubsuit$}\footnotetext{These authors contributed equally to this work.}
    \renewcommand{\thefootnote}{\arabic{footnote}}
    \setcounter{footnote}{0}
\section{Introduction}\label{sec:Intro}
In the original BB84 protocol \cite{bb84}, it is assumed that sender, Alice, sends out single-photon states. However, true single-photon sources are not accessible with today's technology. In actual implementations, phase-randomized weak coherent pulse (WCP) sources are used as they mostly emit single-photon signals with a relatively small probability of emitting multi-photon signals. However, the use of WCP sources opens the door for eavesdroppers. In the photon number splitting (PNS) attack \cite{PNS, Lütkenhaus_2002_PNS}, an adversary siphons off photons from multiphoton signals, revealing some information about the secret key. This information leakage is greatest in the BB84 protocol as Alice and the receiver, Bob, explicitly announce their basis choices. This results in multiphoton pulses being completely insecure, and as a consequence the long-range performance of BB84 is severely limited. 

One solution to this problem is to use the decoy-state method \cite{decoy,decoy2,decoy3}, where Alice decides the pulse intensities randomly before she encodes and transmits the signal. After Bob measures all signals, she announces her choices of intensities. In this way, the attacks by the adversary on multi-photon signals can be detected by estimating how the photon number statistics have changed through the transmission channel. However, the decoy-state method requires additional hardware to modulate the intensities on a round-by-round basis according to a random sequence, which introduces additional side channels and requires extra effort to close the loopholes \cite{Zapatero_2021}.

Another potential solution is to use modified announcement structures. We make use of two such protocols. The first is the SARG04 protocol \cite{sarg04}, wherein Alice implicitly annouces the basis encodings and requires Bob to combine the announcements and his detection outcomes to deduce Alice's encoding bases, while still using the physical BB84 hardware. We further propose a second, new protocol which we call No Public Announcement of Basis (NPAB) BB84, where we forego basis sifting altogether in key generation rounds\footnote{{To the best of our knowledge, there is no existing literature that mentions NPAB BB84. If any reader is aware of relevant references, please feel free to contact the authors.}}. The different basis announcement structures of these protocols reduces the efficiency of PNS attacks, and enable us to extract secret keys from multi-photon signals, without the need for decoy states. Furthermore, BB84, NPAB BB84, and SARG04 can be implemented using the same BB84 hardware as they all simply require Alice sending computational and conjugate basis states, with Bob actively measures in these bases. In this work, we investigate the the performance of  BB84, NPAB BB84, and SARG04 protocols under lossy and noisy channels with the same device setups to understand the impact of announcement structure on key rate. Our proof framework allows for the extraction of secret keys from multi-photon signals, and we illustrate the resulting performance improvements achieved by explicitly including these contributions in the analysis for NPAB and SARG04. 
This paper is organized as follows. In Sec. \ref{sec:Backgrounds} we detail a generic QKD protocol and we further detail the specific protocols investigated in this paper. In Sec. \ref{sec:Method} we briefly describe our mathematical model for simulating the protocol. In Sec. \ref{sec:Results} we detail the resulting protocol performance under various model assumptions, varying channel loss, channel alignment and error rates. Finally in Sec. \ref{sec:Conclusion} we provide some concluding remarks.

\section{Protocol Descriptions}\label{sec:Backgrounds}
\subsection{General QKD Protocol Description}\label{ssec:Protocols}
We begin by describing a generic, variable length QKD protocol.
\vspace{-3mm}
\subsubsection{Signal Preparation} In each round, Alice decides with probability $p_{\text{Gen}}$ whether the sent signal will be used for a key generation or testing round and store this information in a string. Alice then picks a signal state from a set of $d_A$ states $\{\ket{s_x}\}_{x=1,...,d_A}$ with associated probabilities $\{p_x\}_{x=1,...,d_A}$, which she then encodes into a signal with intensity $\mu$ in phase-randomized WCP, sent to Bob through a quantum channel $\mathcal{E}$. She stores the value of $x$ in a classical register $X$. 
\subsubsection{Measurement Phase} Bob measures the incoming states with positive operator-valued measure (POVM) elements $\{\Gamma_b^B\}_{b=1,...,d_B}$, where the index $b$ runs over all possible measurement outcomes, and stores his information in a classical register $Y$. The signal and measurement phases are repeated $N$ times. 
\subsubsection{Announcement and Testing Phase} Alice and Bob process their data using public announcements through an {authenticated} classical channel and sifting their resulting bits. In the subset $m=(1-p_\mathrm{Gen})N$ selected testing, Alice and Bob announce their full measurement outcomes to obtain observed statistics $\vec{F}^{\text{obs}}$. The remaining $n=N-m=p_\mathrm{Gen}N$ rounds are used for key generation. Alice and Bob make public announcements $C_A$ and $C_B$, respectively. Announcements include whether a round is used for testing or for generation, and can include detect/no detect, basis choice or sets of possible signal states as further announcements.
\subsubsection{Key Generation and Privacy Amplification} Alice maps her register $X$ to a binary raw key register $Z$, using a key map $g:\mathcal{X}\times \mathcal{C}_A\times \mathcal{C}_B\to Z\cup\{\perp\}$. This function implements key mappings and sifting of signals depending on the previous classical announcements. The two parties then perform error correction, leaking $\delta_{\text{leak}}$ bits of information, and then apply a two-universal hash function to their raw keys in order to generate a common secret key. 
\subsection{The BB84 hardware}
 BB84, NPAB BB84 and SARG04 can be implemented using the same physical hardware, as well as the same \textit{Signal Preparation} (up to different choices of source settings and basis choices) and \textit{Measurement Phase}. We now explain the setup and the parameters Alice and Bob have the freedom to change in this subsection. 


\begin{description}
    \item[Signal Preparation]: Alice can choose the intensity $\mu$ of the phase-randomized WCP before the protocol begins, which fixes the intensity for all rounds in the protocol.  For each round (for both key generation round and test round), she decides to encode in the $Z\ (X)$-basis with probability $p_z\ (p_x)$. In $Z\ (X)$-basis, she randomly choose the encoding $\{\ket{H},\ket{V}\}\ (\{\ket{D},\ket{A}\})$ with equal probability. Since Alice prepares the outgoing states, she has the choice to misalign the outgoing basis states with respect to the two canonical bases by introducing some rotation angle $\theta$.
    \item[Measurement Phase] Bob has an active detection setup. He measures in $Z\ (X)$-basis with the same probability $p_z\ (p_x)$ that Alice choses the basis encoding. While this is a choice, it maximizes the probability of basis match between the two.
    \item [Announcement and Testing Phase]  For testing rounds, Alice and Bob reveal the encodings and measurement outcomes. For key generation rounds, different protocols have different announcement structure as we will explain in the next subsection. 
\end{description}
We make use of polarization encoding $\{\ket{H},\ket{V},\ket{D},\ket{A}\}$ using a phase-randomized WCP source. {Although we pick this specific encoding, we do not expect major change in the following analysis for encoding using other degrees of freedom such as $\{\ket{R},\ket{L}\}$ or time-bin encoding.}

\subsection{Announcement Structure for Specific Protocols}
We now describe the different announcement structures of BB84, NPAB BB84 and SARG04.
\subsubsection{BB84 Protocol}In the BB84 protocol, Alice randomly decides on sending a bit in either the computational ($\{\ket{H},\ket{V}\}$) or conjugate ($\{\ket{D},\ket{A}\}$) basis. The bit outcome and basis encoding are stored in strings $X\in\{0,1\}^N$ and $X'\in\{0,1\}^N$, respectively. These states are sent through an insecure quantum channel $\mathcal{E}$. Bob uses a randomly generated string $Y'\in\{0,1\}^N$ to decide the basis in which he will measure the incoming signal. After all signals are measured, Alice announces the basis encodings $X'$ and Bob announces measurement bases $Y'$, and they discard any signals with basis mismatch. An extra announcement is made to discard signals where Bob measured vacuum. A subset with $m$ bits is used for {testing} between the two parties. {Alice performs key map on the remaining $n$ bits to obtain the raw key}.
\subsubsection{No Public Announcement of Basis (NPAB) BB84}
NPAB BB84 differs from regular BB84 in the announcement and testing phase and is used to strengthen BB84 against PNS attacks by limiting public basis announcements. In this phase, Alice only reveals basis encodings and outcomes for testing signals. The only announcements made in the key generation rounds are to discard vacuum signals. The remaining steps occur as in the case of BB84 to extract a secret key. This protocol differs from previous reduced-public-announcement BB84 variants, such as the Hwang Protocol \cite{HWANG1998489,HwangImproved}, in that no secret information is required to be shared between the two parties ahead of time. 
\begin{figure}
    \centering
    \includegraphics[width = 0.45\textwidth]{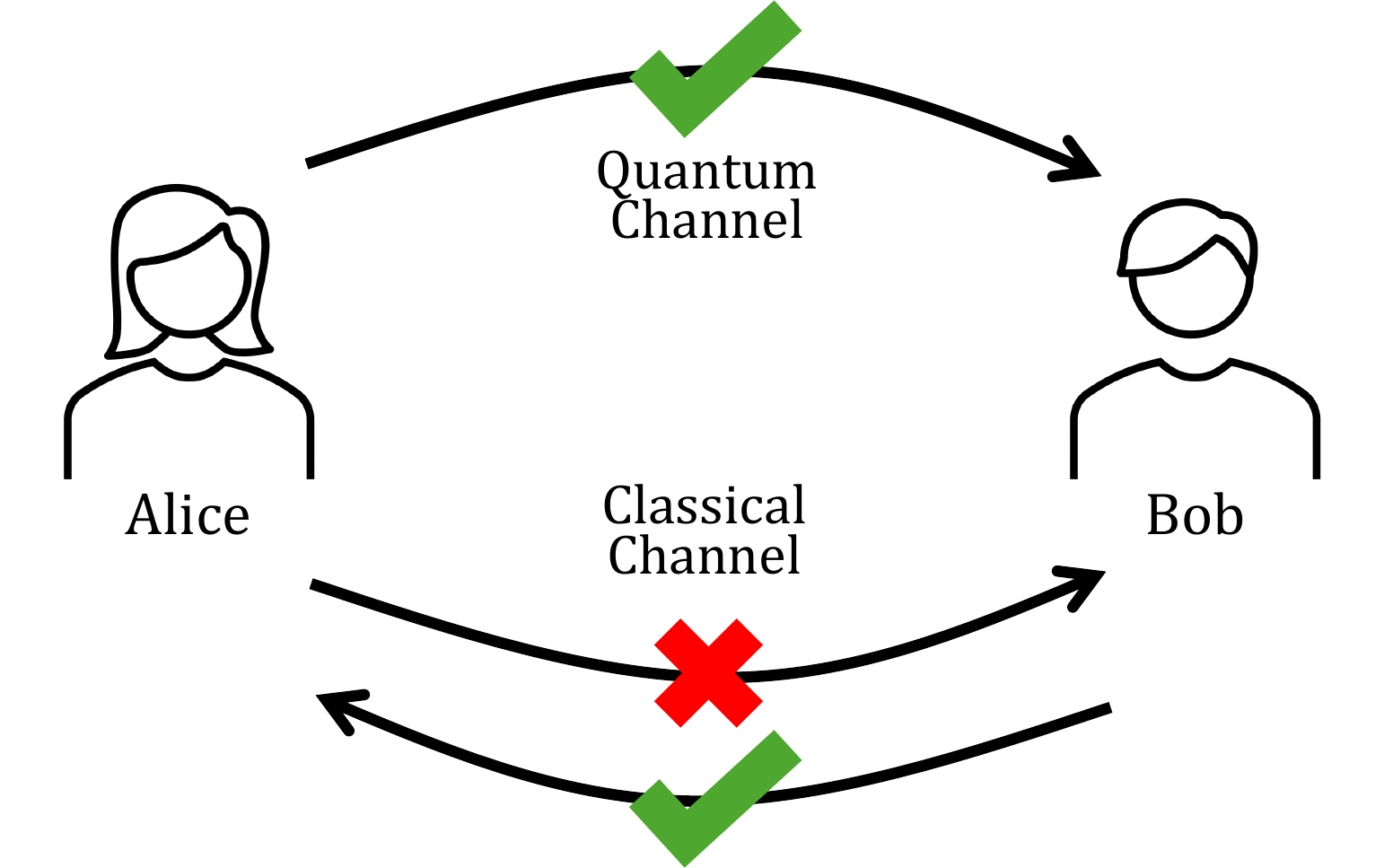}
    \caption{In NPAB BB84, the only sifting done is Bob announcing no click/click/multi-click.}
    \label{fig:npabStates}
\end{figure}
\subsubsection{SARG04}
SARG04 also differs from regular BB84 in the classical announcement phase \cite{sarg04}. During the quantum communication rounds, Alice still sends a polarized photon. We denote this state to be $\ket{s_x} \in \{\ket{H},\ket{V},\ket{D},\ket{A}\}$. Rather than announce polarization encodings, Alice randomly picks one of the two states that are not orthogonal to $\ket{s_x}$. We call this new state $\ket{s'_x}$. She then announces the set $\{\ket{s_x},\ket{s_x'}\}$. If Bob has measured a state orthogonal to $\ket{s_x'}$, he can conclude the correct signal is $\ket{s_x}$. Otherwise, he discards the round. For example, in one particular round, Alice sends an $\ket{H}$ state. In the announcement phase, she can then announces the pair $\{\ket{H},\ket{D}\}$. If Bob measures $\ket{A}$, he can deduce that Alice has sent the $\ket{H}$ state and this round is kept for key generation. Otherwise, he discards it. The remaining steps occur as in the case of BB84 and NPAB BB84 except that for SARG04, Alice maps $\{\ket{H},\ket{V}\}$ to key bit 0 and maps $\{\ket{A},\ket{D}\}$ to key bit 1.
\section{Methods}\label{sec:Method}
In this section, we describe the essential methods we use to numerically calculate key rates. The full details are described in Appendix \ref{sec:Appendix1}.

To numerically represent {Bob's measurement, we use a post-processing map that randomly assigns double click events to one of the outcomes on Bob's side and Bob adopts the qubit and vacuum measurement. This process, referred to as squashing, is required by our proof technique in order to simulate Bob's measurements as qubits \cite{simpleSquashing}. 
The asymptotic, independent and identical distributed (IID) key rate formula \cite[Eq.(50)-Eq.(54)]{ winickReliableNumericalKey2018} is
\begin{align}
    R_{\infty} &= \min_{\rho_{AB}\in \mathcal{S}} D\left(\mathcal{G}(\rho_{AB})||\mathcal{Z}(\mathcal{G}(\rho_{AB}))\right)-\fec\delta_{\text{leak}}
    \\
    &=: \min_{\rho_{AB}\in \mathcal{S}} f(\rho_{AB})-\fec\delta_{\text{leak}},
\end{align}
where $\mathcal{S}$ is a predetermined feasible set constraining channel behaviour, $D(\cdot||\cdot)$ is the quantum relative entropy, $\mathcal{G}$ and $\mathcal{Z}$ {are maps that} model measurements, sifting and key maps, and $\delta_{\text{leak}}$ is the cost for classical error correction with weighting $f_\mathrm{EC}$.

When we consider finite-size effect, given a variable-length protocol with  $\varepsilon_{\text{sec}}=\varepsilon_{EV}+\varepsilon_{PA}+\varepsilon_{AT}$ security,{ where the $\varepsilon's$ are security parameters for error verification ($\varepsilon_{EV}$), privacy amplification ($\varepsilon_{PA}$) and acceptance testing ($\varepsilon_{AT}$)}. Using the post-selection security proof technique, the key length can be shown to be lower bounded against coherent attacks by (\cite{variable_length}, Eq. 49 of \cite{LarsDevAdaptive} with $t=0$)
\begin{align}
    \ell_{\text{finite}} &\geq n_{\text{sift}} \min_{\rho_{AB}\in{\mathcal{V}(\vec{F}^{\text{obs}})}}\frac{1}{P(\text{sift} \wedge \text{gen})}f(\rho_{AB}) -\fec\delta_{\text{leak}}\notag \\
    &-\sqrt{n_{\text{sift}}}(\alpha-1)\log^2(\dim A+1)-\log_2\left(\frac{2}{\varepsilon_{EV}}\right)\notag\\
    &-\frac{\alpha}{\alpha-1}\left(\log(\frac{1}{2\varepsilon_{PA}})+\frac{2}{\alpha}\right),
\end{align}
where, $N$ is the total number of signals, $n_{\text{sift}}$ is the number of signals surviving sifting, $\mathcal{V}(\vec{F}^{\text{obs}})$ is the constraint set generated by the observations and $\alpha$ is the Rényi entropy parameter. 

\section{Results}\label{sec:Results}
    We present an overview the numerical results. We make use of the OpenQKDSecurity software package \cite{softwareZenodo} to simulate the channels and compute key rates. Our findings can be summarized as follows:
\begin{description}
    \item[Misalignment] The case where Alice and Bob's devices are perfectly aligned is not optimal for NPAB BB84. In other words, the key rate performance is better when we intentionally misalign Alice and Bob's device by some angle $\theta$, in order to maximize the probability of matching bits, rather than states.
    \item[Multiphoton Contribution] For NPAB BB84 and SARG04, the use of numerical proof techniques allows us to extract secret information from multi-photon signals, which are typically assumed to be insecure in BB84. We show that allowing Alice and Bob to distill key bits from 2-photon and 3-photon signals does indeed improve secret key rates. 
    \item[Channel Parameter Resilience]We demonstrate that the three protocols have different regimes in which they show their respective advantages. We consider asymptotic and finite key rates for lossy, misaligned and depolarizing channels.
\end{description}
The following are device and proof technique parameters considered in our simulation. We specify if this parameter is optimized for given channel settings. Note that we do not specify channel parameters here. We introduce a new parameter $K$ labelling the photon number cutoff in our simulation, where we assume all photon pulses including more than $K$ photons are completely leaked to the adversary. We refer to Sec. \ref{sec:multiPhotonSignals} for more details.
\begin{center}
\begin{tabular}{ |c|c|c| } 
\hline
Parameter & Symbol &  Optimized vs. Set  \\
\hline
Mean Photon Number & $\mu$ & Optimized
\\
Device Misalignment & $\theta$ & Set
\\
Photon Number Cutoff & $K$ & Set
\\
Security Parameters & $\varepsilon_i$ & Set
\\
\hline
\end{tabular}
\end{center} 
\subsection{Misalignment Angle $\theta^{opt}$ for NPAB BB84}
One possible source of error in implementations of QKD protocols is physical misalignment in the quantum channel. In BB84 or any protocol with basis announcements, the optimal angles are $\theta=0$ and $\pi/2$, as any non-trivial angle introduces some error in distinguishing between encoded key bits. NPAB BB84 requires optimizing the probability of measuring the correct encoded basis without the condition that Alice and Bob are sure of bases matching. Key rates can be increased by rotating detectors by some non-zero angle $\theta^*$. In the qubit case, this value of $\theta^{*}$ is found by maximizing
\begin{align*}
        P(\text{Alice sends $k$ $\wedge$ Bob measures $k$})\\=\Tr\left(\Pi_k\mathds{1}_A\otimes\Phi_B^\theta(\rho_{AB|k})\right),
\end{align*}
where $\rho_{AB|k}$ is the mixed state shared between Alice and Bob conditioned on key bit $k$, $\Pi_k$ is the sum of POVM elements corresponding to bit $k$ being sent and measured and $\Phi^\theta$ is a unitary rotation about the $Y$ axis $\Phi^{\theta}(\rho) =e^{-i\theta\vec{a^{\dagger}}\sigma_Y\vec{a}}\rho e^{i\theta\vec{a^{\dagger}}\sigma_Y\vec{a}}$,
where $\sigma_Y$ is the $Y$ Pauli matrix, $\theta$ is the physical misalignment angle \cite{optics} and $\vec{a^{\dagger}} = [\hat{a}_H^{\dagger} $  
  $\hat{a}_V^{\dagger}]$. The POVM element $\Pi_k$ comes from the fact that in NPAB BB84 the joint POVM elements are coarse grained so that successful key generation rounds rely on bits matching, rather than states matching. In practice, $\theta^{*}$ is roughly a weighted average of the physical angles between states used to encode $k$, with small corrections to minimize double click events resulting from multi-photon signals. To maximize protocol performance, we propose that this $\theta^{\text{opt}}$ misalignment be thought of as part of the protocol setup, and any physical misalignment be deviations from this optimal angle, i.e. $\theta^{\text{protocol}}=\theta^{*}\pm\theta^{\text{error}}$. 
For the remainder of the paper, NPAB BB84 results will be simulated with an optimized non-zero misalignment angle $\theta^{*}$ in order to boost protocol performance (Fig. \ref{fig:npabStates}). 
In both the asymptotic and finite-size regimes, NPAB BB84 is more resilient to physical misalignment and is capable of outperforming both BB84 and SARG04 (Fig. \ref{fig:misalignAsym}, Fig. \ref{fig:misalign12}, Fig. \ref{fig:misalign9}). 
\vspace{-3mm}
\begin{figure}[H]
    \centering
    \includegraphics[width = 0.45\textwidth]{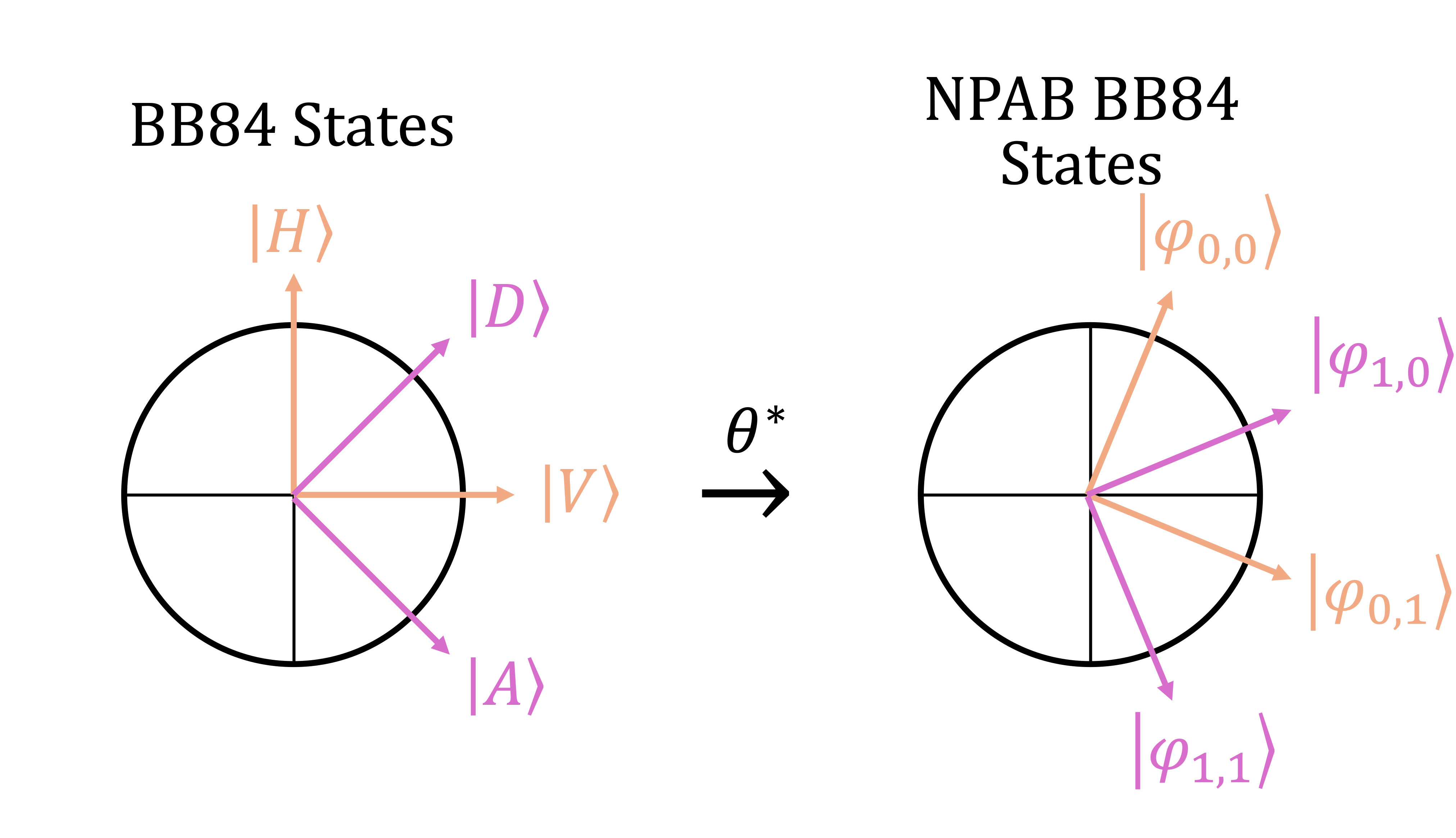}
    \caption{Optimized NPAB BB84 protocol states are shifted relative to bases.}
    \label{fig:npabStates}
\end{figure}
\begin{figure}[H]
    \centering
    \includegraphics[width = 0.52\textwidth]{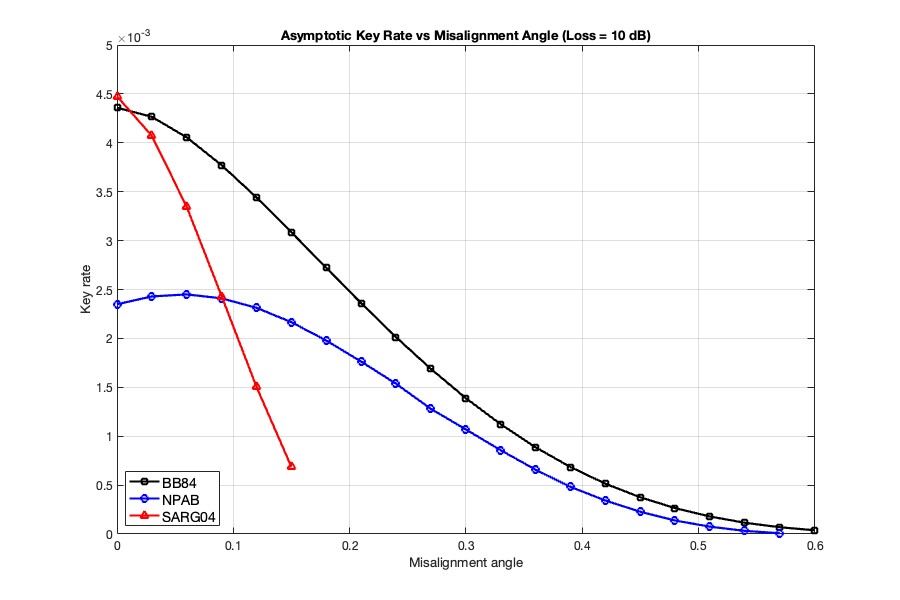}
    \caption{Performance of BB84, SARG04 and NPAB BB84 at 10 dB channel loss when introducing physical misalignment $\theta$ between detectors in the limit $N\rightarrow\infty$. The signal intensity is optimized.}
    \label{fig:misalignAsym}
\end{figure}
\subsection{Inclusion of Multiphoton Signals}\label{sec:multiPhotonSignals}
For NPAB and SARG04, multi-photon signals contribute to the key rates due to bases not being announced and we can choose to include those signals in our analysis. Key rates contribution from signals with photon number greater than 4 is small (Fig. \ref{fig:ncut}). Since the increase in key rates from $K=3$ to $K=4$ is relatively small, and it takes much more computer resources to calculate key rates when increasing $K$, we include $K=3$ number of photons in key rates calculation for NPAB and SARG04 and flag the remaining signals. For mathematical details, see Sec. \ref{apd:squash}.
   
    \begin{figure}
   \centering
    \includegraphics[width=0.52\textwidth]{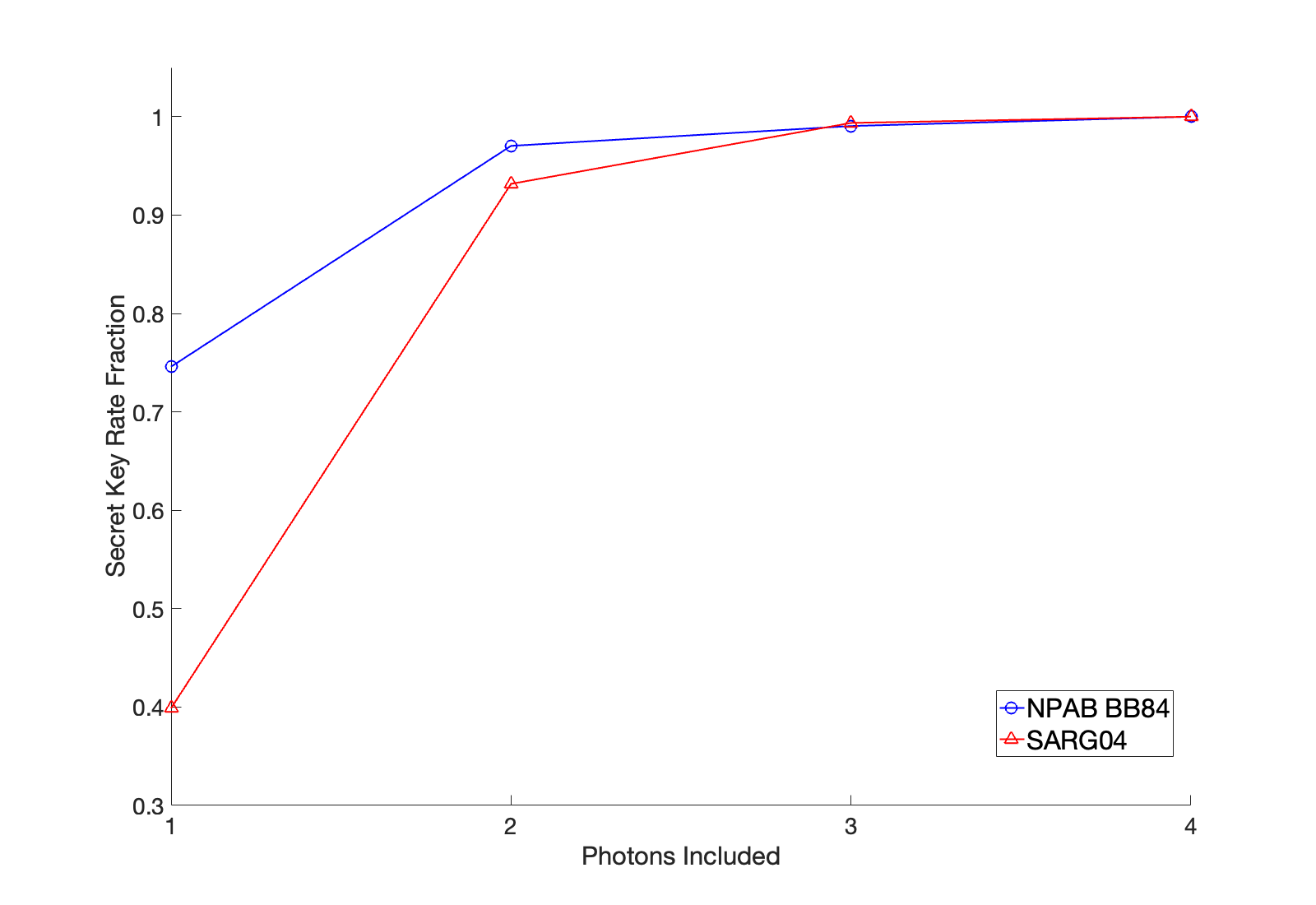}
    \caption{
    Key rate versus number of photons used for key calculation in a given pulse (the photon cutoff $K$). Protocols simulated in the asymptotic limit $N\rightarrow\infty$ with 5 dB quantum channel loss and we optimize the choice of intensity of the phase-randomized WCP (mean photon number $\mu$).}
   \label{fig:ncut}
    
\end{figure}
\subsection{Asymptotic Key Rates}
\vspace{-2mm}
In this section, we explore the performance of BB84, SARG04 and NPAB BB84 in the asymptotic regime using the framework in \cite{asymptotics} (See \ref{apd:Asym}, \ref{apd:Loss}) in which we assume Alice sends infinitely many signals allowing for exact bounds on channel statistics. 

We first investigate the performance of a loss-only channel. We find that SARG04 is more resilient to high loss, allowing for higher key rates in that regime. These results agree with a previous analytic study on SARG04 and BB84 \cite{Renner}. In addition, NPAB BB84 does not show any advantage against the other two protocols (Fig. \ref{fig:rAsym}). Note that the announcement structure of SARG04 allows for a higher signal intensity and mean photon number, with $\mu^{\text{opt}}_{\text{SARG04}}\sim 10^{-1}$ between $0$ and $35$ dB (Fig. \ref{fig:muAsym}). 
\begin{figure}[H]
    \centering
    \includegraphics[width=0.52\textwidth]{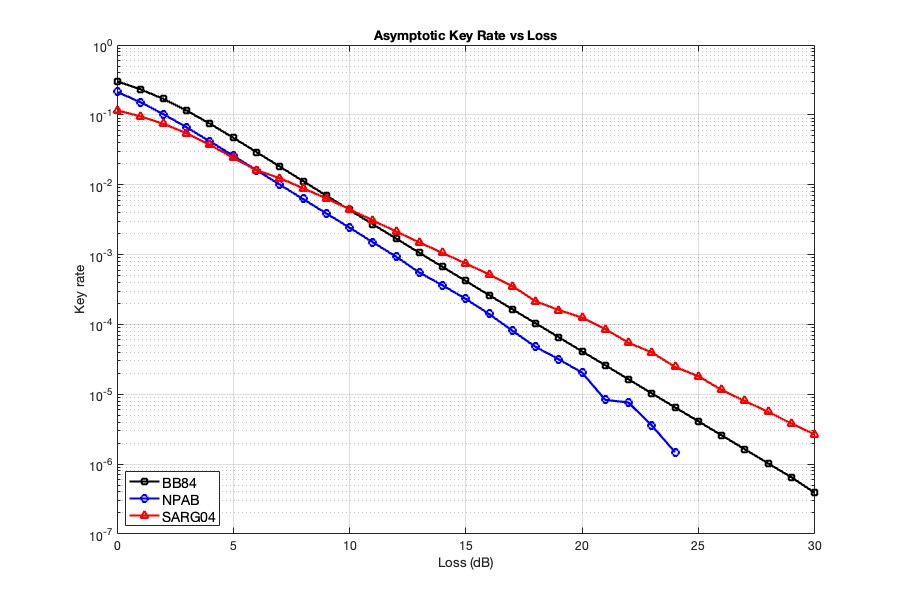}
    \caption{Secret key rate per signal in a loss-only channel for protocols in the asymptotic limit $N \rightarrow\infty$ total signals sent. The signal intensity is optimized. We take $\text{loss} = -10\log_{10}\eta$ in dB.}
    \label{fig:rAsym}
\end{figure}
\begin{figure}[H]
    \centering
    \includegraphics[width=0.52\textwidth]{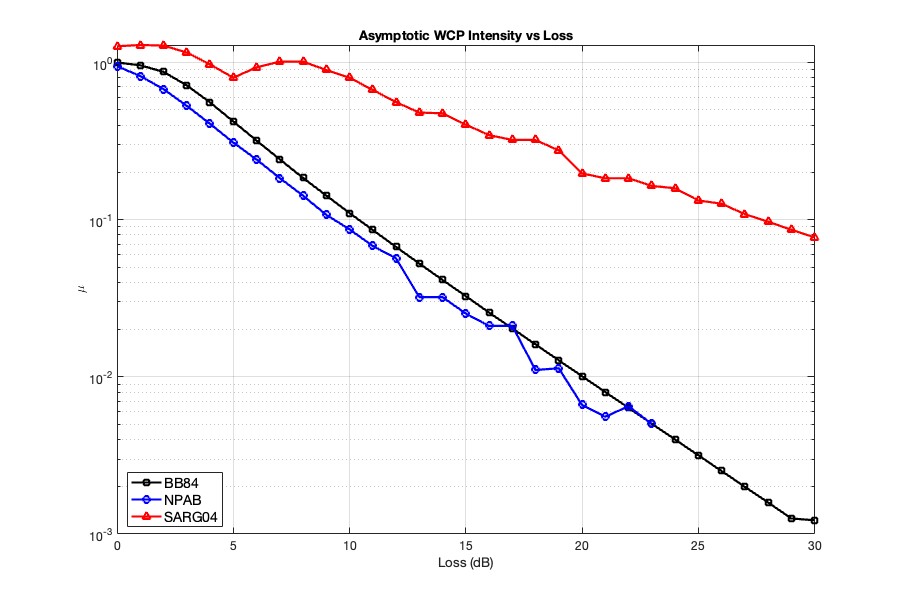}
    \caption{Optimal pulse intensity (mean photon number) versus loss in dB in a loss-only channel for asymptotic key rates.}
    \label{fig:muAsym}
\end{figure}
Next, we introduce depolarization error into the channel (\ref{apd:Depol}). The advantage of SARG04 vanishes quickly as error increases due to its announcement structure. This also agrees with previous results indicating SARG04 is not robust against depolarization error \cite{Renner}. Again, however, NPAB BB84 shows no significant advantage compared to BB84, though it does perform better than SARG04 (Fig. \ref{fig:depolAsym}). 
\begin{figure}[H]
    \centering
    \includegraphics[width=0.52\textwidth]{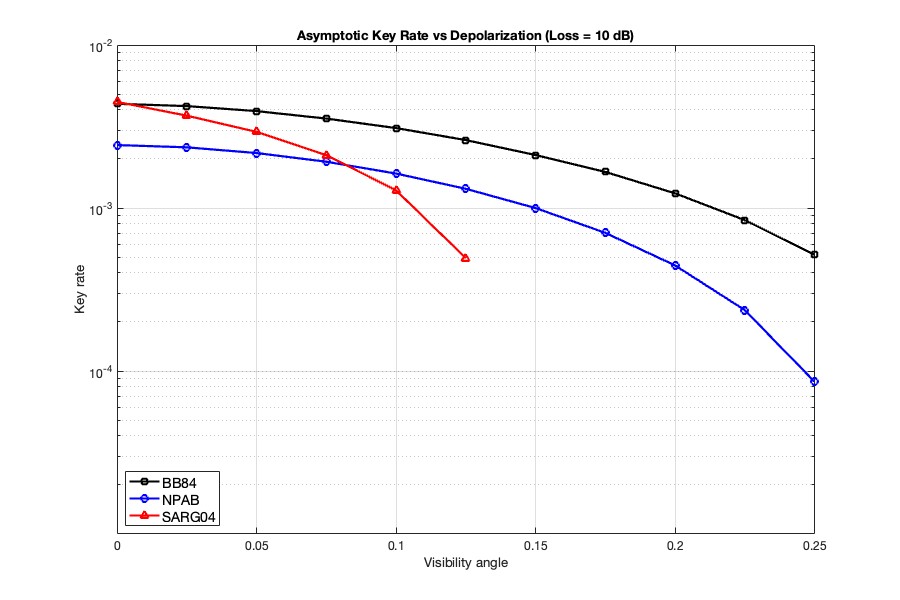}
    \caption{Secret key rate per signal in a quantum channel with depolarizing noise and fixed 10dB loss for protocols in the asymptotic limit $N \rightarrow\infty$ total signals sent.}
    \label{fig:depolAsym}
\end{figure}
\subsection{Finite-size Key Rates}
We repeat the previous section's investigation while introducing finite-size effects (See \ref{apd:finite}). We specifically investigate the case for $N=10^{12}$, $N=10^9$ and $N=10^6$ total signals, corresponding to 1000 seconds, 1 second and 1 millisecond with a 1 GHz clock rate respectively, representing current and near-future technical capabilities respectively \cite{vajner2022quantum}. 
As before, we begin with investigating the behavior of the protocols in a loss-only channel. As was the case in the asymptotic formulation, a loss-only protocol implementation with $N=10^{12}$ total signals and $p_{\text{Gen}}=0.85$ demonstrates that SARG04 is much more resilient to loss, with the relative scaling between the three protocols being similar (Fig. \ref{fig:r1e12}). 
Limiting the total number of signals to $N=10^9$ and $N=10^6$ introduces much more prevalent finite-size effects. We observe that the general behaviour is similar in the asymptotic case, with SARG04 having stronger performance with high loss. Due to the finite-size effects, expected key rates die out at a quicker rate (Fig. \ref{fig:r1e9}, Fig. \ref{fig:r1e6}). Compared to $N=10^{12}$ signals, which behaves close to the asymptotic limit, key rates cannot tolerate as much loss and the protocols are unable to produce key rates past 15 to 20 dB. 
\begin{figure}[H]
    \centering
    \includegraphics[width=0.52\textwidth,height=0.21\textheight]{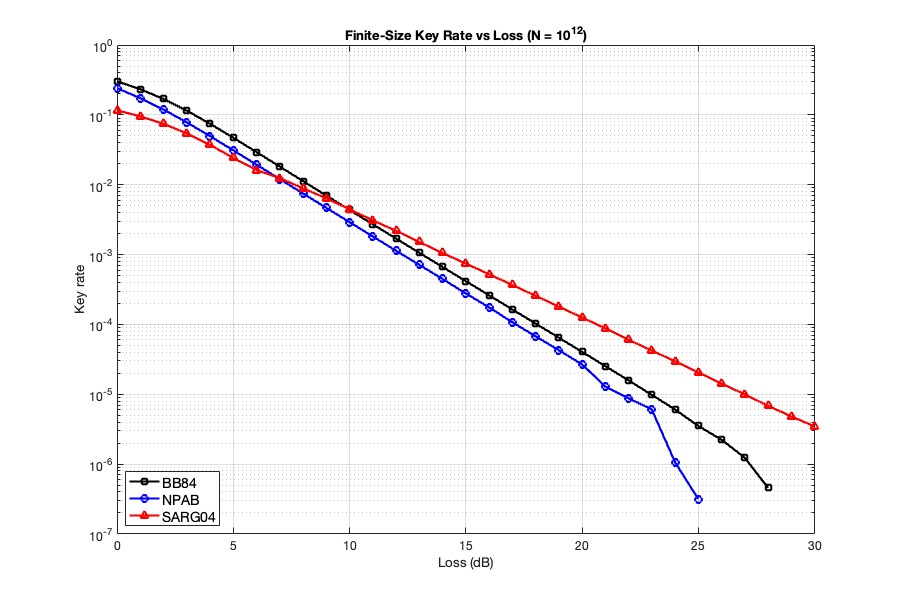}
    \caption{Secret key rate per signal for a loss-only quantum channel for protocols with $N = 10^{12}$ total signals sent. The signal intensity is optimized.}
    \label{fig:r1e12}
\end{figure}

\begin{figure}[H]
    \centering
    \includegraphics[width=0.52\textwidth,height=0.21\textheight]{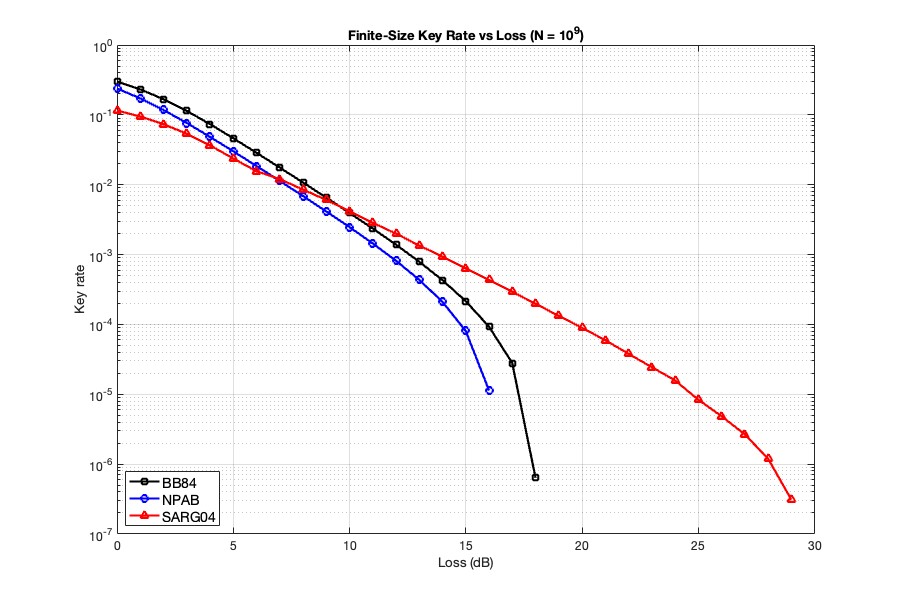}
    \caption{Secret key rate per signal in a loss-only quantum channel for protocols with $N = 10^{9}$ total signals sent. The signal intensity is optimized.}
    \label{fig:r1e9}
\end{figure}

\begin{figure}[H]
    \centering
    \includegraphics[width=0.52\textwidth,height=0.21\textheight]{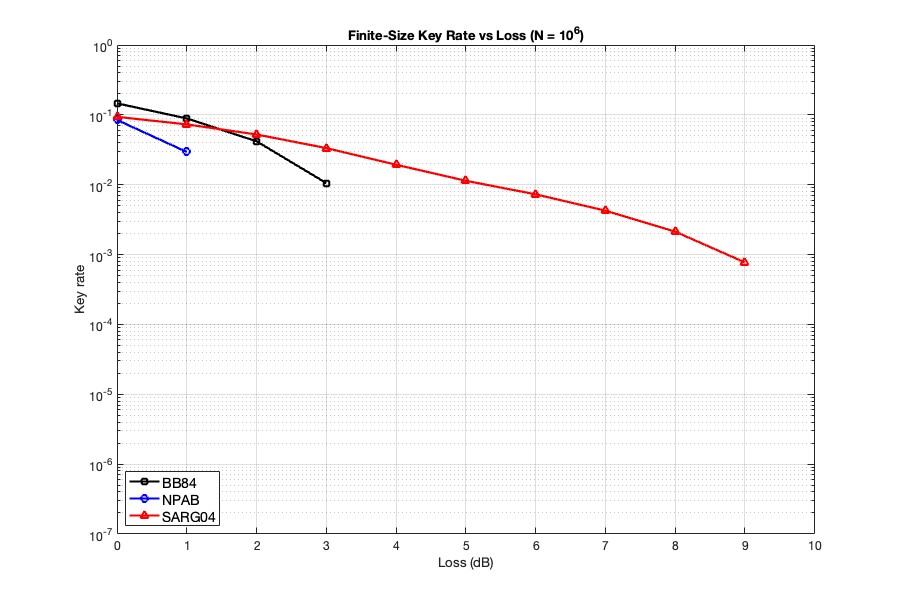}
    \caption{Secret key rate per signal in a loss-only quantum channel for protocols with $N = 10^{6}$ total signals sent. The signal intensity is optimized.}
    \label{fig:r1e6}
\end{figure}
\begin{figure}[H]
    \centering
    \includegraphics[width=0.52\textwidth,height=0.23\textheight]{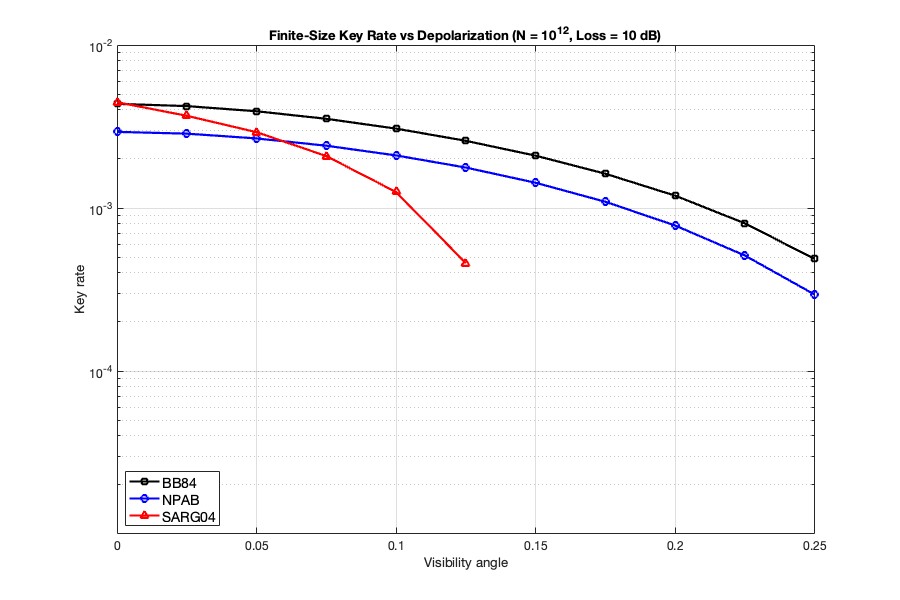}
    \caption{Secret key rate per signal in a quantum channel with depolarizing noise and fixed 10dB loss for protocols in the case of $N =10^{12}$ total signals sent. }
    \label{fig:depol1e12}
\end{figure}
\vspace{-2cm}
\begin{figure}[H]
    \centering
    \includegraphics[width=0.52\textwidth,height=0.23\textheight]{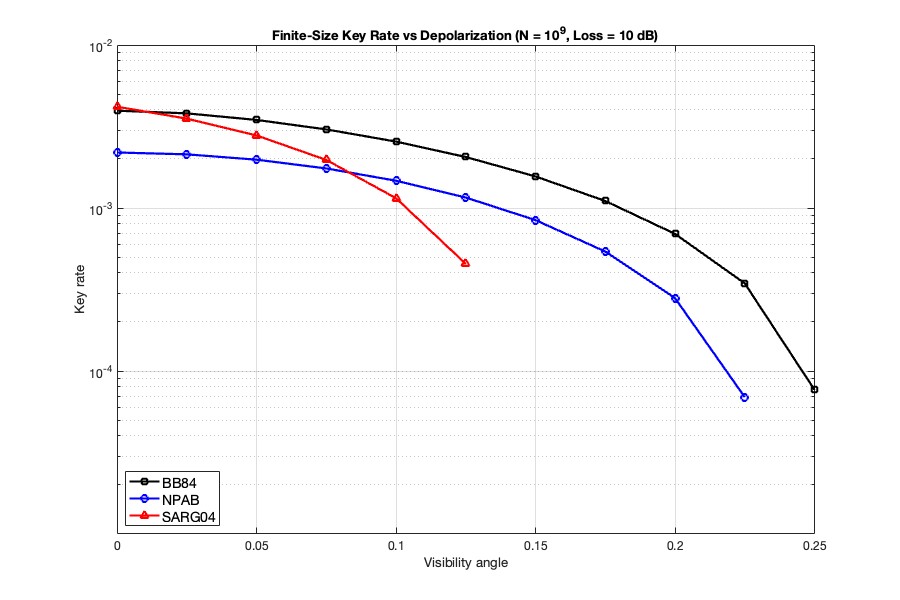}
    \caption{Secret key rate per signal in a quantum channel with depolarizing noise and fixed 10dB loss for protocols in the case of $N =10^{9}$ total signals sent. }
    \label{fig:depol1e12}
\end{figure}
Introducing misalignment and depolarizing noise into the channel leads to similar behavior as before in the asymptotic regime. We observe that BB84 and NPAB BB84 behave similarly. SARG04 is more susceptible to depolarization errors, while BB84 and NPAB BB84 manage higher key rates with error (Fig. \ref{fig:depol1e12}). We do not include calculations for $N=10^6$ for misalignment and depolarization because most of the key rates vanish.
\begin{figure}[H]
    \centering
    \includegraphics[width=0.52\textwidth]{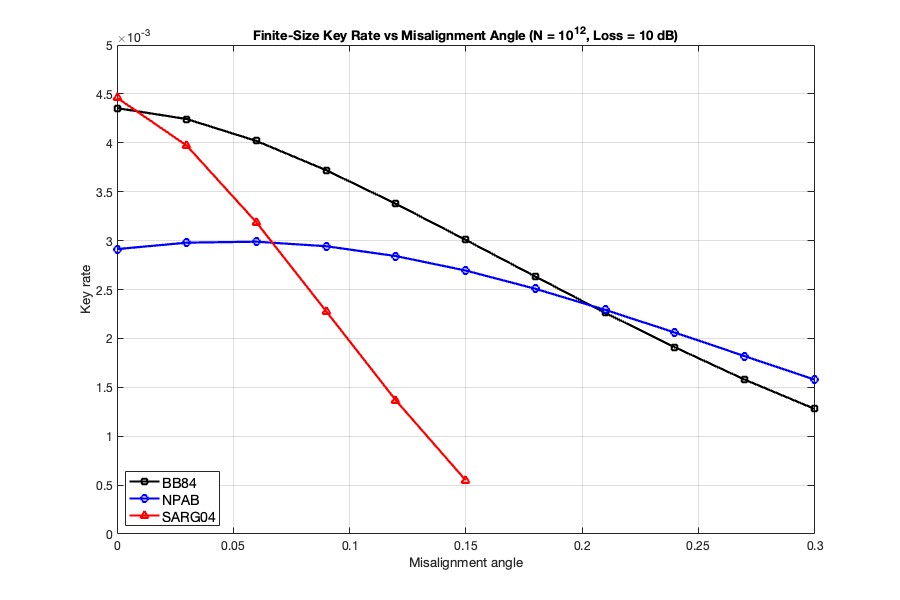}
    \caption{Secret key rate per signal in a quantum channel with physical misalignment and fixed 10dB loss for protocols in the case of$N =10^{12}$ total signals sent.}
    \label{fig:misalign12}
\end{figure}
\begin{figure}[H]
    \centering
    \includegraphics[width=0.52\textwidth]{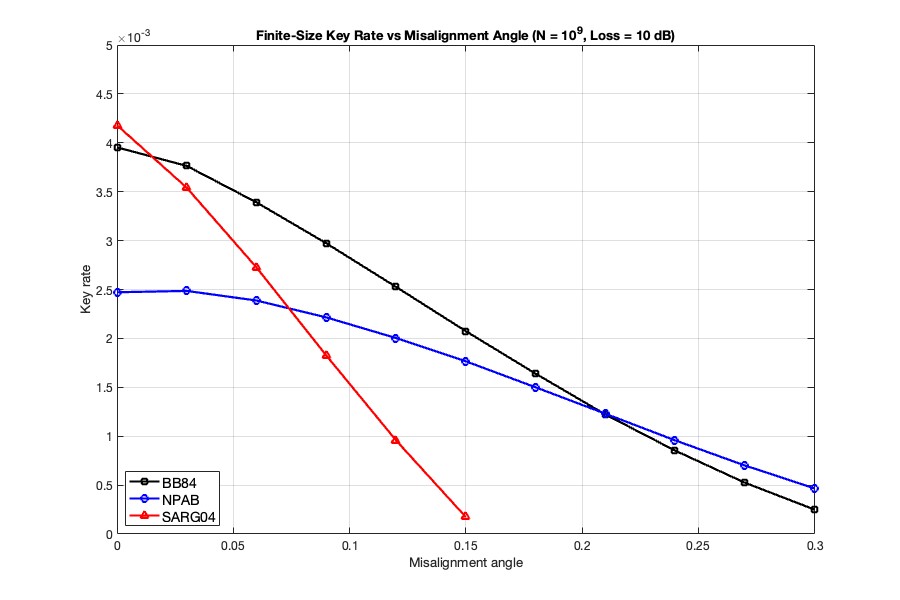}
    \caption{Secret key rate per signal in a quantum channel with physical misalignment and fixed 10dB loss for protocols in the case of $N =10^{9}$ total signals sent.}
    \label{fig:misalign9}
\end{figure}
\section{Conclusion}\label{sec:Conclusion}
{We use numerical security analysis to characterize the performance of three quantum key distribution protocols: BB84, NPAB BB84 and SARG04 with phase-randomized WCP implementations without using the decoy method.} In general, in the case where we do not have decoy states, BB84 is still more robust than NPAB BB84 and SARG04 when considerable loss and noise are present. We have also shown that NPAB BB84 outperforms the other protocols when we consider physical misalignment and that SARG04 can produce more secret key when in a high-loss but noiseless channel.

Further points of interest could include comparing a larger variety and complexity of announcement structures in order to best take advantage of each protocol's specific strengths, as well as changing the physical protocol setups. 
\section{Code Availability}
The code used in this paper is available at\\ \url{https://openqkdsecurity.wordpress.com/repositories-for-publications/}.
\section*{Acknowledgements}
\setlength{\parskip}{3pt}
{This research is supported NSERC (through the NSERC Discovery program and NSERC Alliance project) and QUINT (Qeyssat User INvestigation Team)}
The authors would like to thank John Burniston for his help in implementing the protocols using OpenQKDSecuirity. They further extend thanks to Lars Kamin and Devashish Tupkary for their knowledge of the theoretical aspect of QKD protocols.  
\bibliography{references}
\appendix
\section{Mathematical Model}\label{sec:Appendix1}
\subsection{State Preperation}
We make use of the source replacement scheme to model a prepare-and-measure protocol with an entanglement based one \cite{source_repalcement,sourcerep2}. Each round in a prepare-and-measure protocol consists of Alice sending some signal state $\ket{s_x}$ with probability $p_x$. Through the source-replacement scheme, this is equivalent to Alice preparing and sending the entangled state
\begin{align}
    \ket{\psi}_{AA'}=\sum_{a}\sqrt{p_x}\ket{x}_A\ket{s_x}_{A'}
\end{align}
where the set $\{\ket{x}_A\}_{x=1,..,d_A}$ is an orthonormal basis for Alice's register of output states and the register $A'$ is the signal system before the action of the channel. 
The shield system \cite{shield}, which is necessary if Alice uses a phase randomized source, allows us to write a purification of phase randomized Fock states as
\begin{align}
    \ket{\psi}_{A_SAA'} = \sum_{n=0}^\infty\sqrt{p_n}\ket{n}_{A_S}\sum_x\sqrt{p_{x|n}}\ket{x}_A\ket{s_{x,n}}_{A'}
\end{align}
where $p_n\equiv P_\mu(n)=e^{-\mu}\mu^n/n!$ is the Poisson distribution for the photon number $n\in\N_0$, $p_{x|n}=P(x|n)$ and the signal states are defined in terms of the creation operators \cite{Norbert_thesis} as
\begin{align}
    \ket{s_{x,n}} = \frac{1}{\sqrt{n!}}(a^{\dagger}_{h(x)})^n\ket{0},
\end{align}
where $h$ is a bijective map $h:\{1,2,3,4\}\to\{H,V,D,A\}$.
Finally, Alice sends the signal system through a quantum channel $\mathcal{E}$, after which Alice and Bob share the state
\begin{align}
    \rho_{A_SAB}=\mathds{1}_{A_SA}\otimes\mathcal{E}_{A'\to B}\left(\ketbra{\psi}{\psi}_{A_SAA'}\right). \label{eq:shared}
\end{align}

\subsection{Tagging Source and Squashing Model}\label{apd:squash}
In order to numerically represent this system with finite matrices, we model the incoming phase randomized state up to some photon number cutoff \cite{tag}. {In our calculation, we pick $K = 3$.} Past this cutoff, we tag those states with orthogonal states $\{\ket{f_x}\}_{x=1,...,d_A}$, leaving Eve with full knowledge of tagged states. The shared density matrix is then
\begin{align}
        \rho_{A_SAA'} = \ketbra{\Psi}_{A_SAA'}
\end{align}
where 
\begin{align}
    \ket{\Psi}_{A_SAA'} = \sum_{n=0}^K\sqrt{p_n}\ket{n}_{A_S}\otimes \sum_{x}\sqrt{p_{x|n}}\ket{x}_A\otimes\ket{s_{x,n}}_{A'}\notag\\
    +\ p_{n>K}\ket{K+1}_{A_S}\otimes\sum_{x}\sqrt{p_{x|K+1}}\ket{x}_A\otimes\ket{f_x}_{A'}.
\end{align}
On Bob's side, we use the squashing map {associated with the post-processing which }randomly assigns double click events to one of the detection outcomes. This squashing model also allows us to represent Bob's measurements in the qubit and vacuum subspace \cite{simpleSquashing}.
\subsection{Measurements}
Following the framework of \cite{asymptotics, winickReliableNumericalKey2018}, in the prepare and measure implementation Alice's choice of signal state is equivalent to performing a measurement on $\rho_{A_SA}$ with POVM defined by elements
\begin{align}
    \Gamma^A_x\in\left\{\sum_{n=0}^{K+1}\ketbra{n}{n}\otimes\ketbra{x}{x}\right\}_{x= 1,...,d_A}.
\end{align}
Accordingly, Bob's measurements are represented as basic qubit measurements, with POVM elements
\begin{align}
    \Gamma^B_y\in \left\{\ketbra{y}{y}\right\}_{y=H,V,D,A,\perp}.
\end{align}
\subsection{Kraus Operators and Key Map} \label{apd:kraus}
In the framework of \cite{winickReliableNumericalKey2018}, we use a set of Kraus operators $K_{\alpha,\beta}:=K^A_\alpha\otimes K^B_\beta$ to incorporate Alice and Bob's announcements and the key map, where the indices $\alpha,\beta$ run over Alice and Bob's possible announcements. Let $R$ be the key register, $X$ a register of Alice's measurement outcomes, $Y$ a register of Bob's measurement outcomes, $C=C_A\times C_B$ a register of the joint announcements, and let $\mathcal{X}_{\alpha}$ bet the set of Alice's measurement outcomes associated with announcement $\alpha$, and $\mathcal{Y}_{\beta}$ be a set of measurement Bob's measurement outcomes with announcement $\beta$. Let $g:\mathcal{X}\times \mathcal{C}\to Z^n$ be the key map, which takes in announcements and Alice's measurement outcome (post-sifting, stored in alphabets $\mathcal{C}$ and $\mathcal{X}$ respectively) and outputs a binary raw key. $\Gamma^A_{\alpha,x}$ and $\Gamma^B_{\beta,y}$ are the POVM elements corresponding to the announcement. Then the joint Kraus operator is defined as

\begin{align}
    K_{\alpha,\beta} = \notag\Pi^{\text{sift}}_{RXYABC}\sum_{\substack{x\in\mathcal{X}_\alpha \\y\in\mathcal{Y_\beta}}}\ket{g(x,\alpha,\beta)}_R\otimes\ket{x}_{X}\otimes\ket{y}_Y
    \\
    \otimes\sqrt{\Gamma^{A}_{\alpha,x}}\otimes\sqrt{\Gamma^B_{\beta,y}}\otimes\ket{\alpha,\beta}_C
\end{align}
where 
\begin{align}
    \Pi^{\text{sift}}_{RXYABC} = \mathds{1}_{RXYAB}\otimes\Pi^{\text{sift}}_C.
\end{align}
The map $\Pi^{\text{sift}}_{RXYABC}$ performs sifting of signals based on $C$, sifting out testing, vacuum and, in the case of BB84 and NPAB BB84, basis-mismatched signals.
We are left with surviving Kraus operators, explicitly described in the following sections, which define a map we denote by $\mathcal{G}$.
Furthermore, we define a $\mathcal{Z}$ map is given by
\begin{align}
    \mathcal{Z}(\rho) =\sum_{r} P_r \rho P_r^\dagger,
\end{align}
where 
\begin{align}
    P_r = \ketbra{r}{r}_R \otimes \mathds{1}_{XYABC}.
\end{align}
\subsection{Asymptotic Key Rates}\label{apd:Asym}
We compute asymptotic key rates using the framework developed in \cite{winickReliableNumericalKey2018} and improved in \cite{asymptotics}. In this formulation, key rates formula is
\begin{align}
    R_{\infty} &= \min_{\rho_{AB}\in \mathcal{S}} D\left(\mathcal{G}(\rho_{AB})||\mathcal{Z}(\mathcal{G}(\rho_{AB}))\right)-\fec\delta_{\text{leak}}\notag
    \\
    &=:\min_{\rho_{AB}\in \mathcal{S}}f(\rho_{AB})-\fec\delta_{\text{leak}},
\end{align}
where $\mathcal{S}$ is a set constraining the system's density matrix, $D$ is the quantum relative entropy and the $\mathcal{G}$ and $\mathcal{Z}$ maps model measurements, announcements, sifting and key maps. The parameter $f_{EC}$ is a choice characterizing the efficiency of error correction ($f_{EC}=1$ being the Shannon limit). {In our calculation, we pick $f_{EC} = 1$ for asymptotic key rate and $f_{EC} = 1.2$ for finite-size key rate.}
\subsection{Finite-size and Variable Length Key Rates}\label{apd:finite}
In order to compute key rates in consideration of finite-size effect (i.e. given some finite $N$ signals sent between Alice and Bob), new correction terms are needed \cite{variable_length, LarsDevAdaptive}. We compute variable length key rates assuming the channel behaves as expected, such that the observed statistics $\vec{F}^{\text{obs}}$ are equal to the expected channel output $\overline{F}$. Given a protocol with $\varepsilon_{\text{sec}}=\varepsilon_{EV}+\varepsilon_{PA}+\varepsilon_{AT}$ security, the key length is lower bounded by
\begin{align}
    \ell_{\text{finite}} &\geq\ n_{\text{sift}}\min_{\rho_{AB}\in{\mathcal{V}(\vec{F}^{\text{obs}})}}\frac{1}{P(\text{sift} \wedge \text{gen})_{\rho_{AB}}}f(\rho_{AB}) -\fec\delta_{\text{leak}}\notag \\
    &-\sqrt{n_{\text{sift}}}(\alpha-1)\log^2_2(\dim A+1)-\log_2\left(\frac{2}{\varepsilon_{EC}}\right)\notag\\
    &-\frac{\alpha}{\alpha-1}\left(\log_2\left(\frac{1}{2\varepsilon_{PA}}\right)+\frac{2}{\alpha}\right)\label{finiteKey1}
\end{align}
where $N$ is the total number of signals, $n_{\text{sift}}$ is the number of generation signals surviving sifting, $\mathcal{V}(\vec{F}^{\text{obs}})$ is a constraint set accounting for statistical fluctuations in the finite-size regime, and $\alpha$ is the Rényi entropy order with optimal value
\begin{align}
    \alpha^{\text{opt}} = 1+\sqrt{\frac{{\log(1/\varepsilon_{PA})}}{\log^2(\dim A+1) n_{\text{sift}}}}.
\end{align}
In the formalism of Kamin, L. et al (Lemma 7 \cite{LarsDevAdaptive}), the feasible set is defined as 
\begin{align}
    \mathcal{V}(\vec{F}^{\text{obs}}) = \{\rho\in S_\circ\ \vert\ \Tr_{B}(\rho) = \rho_{A_SA}, \notag
    \\
    \Tr(\Gamma_k\rho)\in\{F^{\text{obs}}_k-\widetilde{\kappa}_k^L, F^{\text{obs}}_k +\widetilde{\kappa}_k^U\}\ \forall k\in\Sigma\}
\end{align}
where $S_\circ$ denotes the set of all density matrices over Alice and Bob's systems, $\rho_{A_SA}$ is the fixed state in Alice's lab, $\Gamma_k=\Gamma^A_x\otimes\Gamma^B_y$ for $k = (x,y)$, $\Sigma=\Sigma_{\text{test}}\cup\{\text{sift}\wedge\text{gen}, \perp\}$ is the set of all possible pairs of announcements in union with the events that Alice and Bob abort a given round or not. For test rounds, we define
\begin{align}
    &\widetilde{\kappa}_k^L = F^{\text{obs}}_k - B\left(\frac{\varepsilon_{\text{AT}}}{2\text{card}(\Sigma)}; N_{\text{test}}F^{\text{obs}}_k, N-N_{\text{test}}F^{\text{obs}}_k + 1 \right),\notag 
    \\
    &\widetilde{\kappa}_k^U = -F^{\text{obs}}_k + \notag \\
    &B\left(1-\frac{\varepsilon_{\text{AT}}}{2\text{card}(\Sigma)}; N_{\text{test}}F^{\text{obs}}_k + 1, N-N_{\text{test}}F^{\text{obs}}_k \right)\label{kappas}
\end{align}
where $\mathrm{card}(\cdot)$ denotes the alphabet's cardinality, $B(p;a,b)$ is the $p^{\text{th}}$ quantile of the beta distribution with shape parameters $a$, $b$. Through this formalism, we also simplify the optimization problem by upper bounding the relative entropy prefactor in \eqref{finiteKey1} by
\begin{align}
    &\min_{\rho_{AB}\in{\mathcal{V}(\vec{F}^{\text{obs}})}}\frac{1}{P(\text{sift} \wedge \text{gen})_{\rho_{AB}}}f(\rho_{AB}) \notag\\
    \leq& \frac{1}{F^{\text{obs}}_{\text{sift}\wedge\text{gen}} + \widetilde{\kappa}_{\text{sift}\wedge\text{gen}}^{U}}\min_{\rho_{AB}\in{\mathcal{V}(\vec{F}^{\text{obs}})}}f(\rho_{AB}),
\end{align}
where ${F}^{\text{obs}}_{\text{sift}\wedge\text{gen}}$ is the observed frequency of bits chosen for generation rounds surviving sifting, and $\widetilde{\kappa}_{\text{sift}\wedge\text{gen}}^{U}$ is defined as in \eqref{kappas}, except with all generation rounds $N_{\text{sift}\wedge\text{gen}}$ instead of $N_{\text{test}}F_k$.
\subsection{Modelling the Channel}
\subsubsection{Loss}\label{apd:Loss}
We simulate loss by attenuating the mean photon number of Alice's prepared state. Given a loss of $\lambda$ (in dB), the channel transmittance is defined as 
\begin{align}
    \eta = 10^{-\lambda/10}.
\end{align}
In order to model lost signals due to a lossy quantum channel, we perform the following map to {the phase-randomized weak coherent pulse.}
\begin{align}
    \frac{1}{2\pi}\int_0^{2\pi}\ketbra{\mu e^{i\theta}}{\mu e^{i\theta}}d\theta \mapsto \frac{1}{2\pi}\int_0^{2\pi}\ketbra{\eta\mu e^{i\theta}}{\eta\mu e^{i\theta}}d\theta,
\end{align}

effectively attenuating the mean photon number from $\mu$ to $\mu'=\eta\mu$ \cite{norbertSecurity2000}.
\subsubsection{Misalignment Angle} \label{apd:misalgn}
In order to model physical rotations, we act on the incoming state $\ket{\psi}$ by a unitary rotation about the $y$ axis as our protocols of interest use $Z$ and $X$ basis polarizations. This yields the map
\begin{align}
    \Phi^{\theta}(\rho) =e^{-i\theta\vec{a^{\dagger}}\sigma_Y\vec{a}}\rho e^{i\theta\vec{a^{\dagger}}\sigma_Y\vec{a}},
\end{align}
where $\sigma_Y$ is the $Y$ Pauli matrix, $\theta$ is the physical misalignment angle \cite{optics} and {$\vec{a^{\dagger}} = [\hat{a}_H^{\dagger} $  
  $\hat{a}_V^{\dagger}]$. }
\subsubsection{Depolarization and Visibility}\label{apd:Depol}
In the qubit case, one can simulate errors caused by misalignment in random directions by depolarization. Specifically, let $\ket{\psi}$ be some general qubit state. We can assume that it has real entries in measurement basis $\{\ket{0},\ket{1}\}$ if we only encode signals in Z and X bases. It then follows that
\begin{align*}
     \rho &= \frac{1}{2}(\Phi^{\theta}(\rho_0)+\Phi^{-\theta}(\rho_0))
     \\
          &= \frac{1}{2}(e^{i\theta\sigma_Y}\rho_0 e^{-i\theta\sigma_Y}+e^{-i\theta\sigma_Y}\rho_0 e^{i\theta\sigma_Y})
     \\
          &=\rho_0 (\cos^2\theta-\sin^2\theta) + \sin^2\theta \mathds{1}
        \\
          &=\rho_0 (1-2\sin^2\theta) + \frac{2\sin^2\theta}{2}\mathds{1},
\end{align*}
where $\rho_0 = \ketbra{\psi}$ and $\Phi^{\theta}$ is defined in \ref{apd:misalgn}. Note that here the rotation is reduced to the rotation on the Bloch sphere for the qubit case.
We obtain the standard depolarization channel on the qubit with depolarization probability $p_{\text{error}} = 2\sin^2\theta$.
Therefore, in order to simulate a multi-photon analogy to visibility and loss of coherence \cite{Renner}, we simulate two separate channel observations with some physical misalignment angle of $\varphi$ and another with physical misalignment of $-\varphi$ to generate observations
\begin{align}
    \vec{F}^{\text{obs}} = \frac{1}{2}\left( \vec{F}_{+\varphi} + \vec{F}_{-\varphi}\right).
\end{align}
In the qubit case, depolarization is characterized by channel visibility $V\in[0,1]$. The probability of introducing an error in a signal is then $p_{\text{error}}=\frac{1-V}{2}$. Here, we inherit this idea and define the angle 
\begin{align}
    \varphi = \frac{1}{2}\arccos(V).
\end{align}
Our construction of $\vec{F}^{\text{obs}}$ reduces to the usual depolarized observations with channel visibility $V$ for qubits.
\section{Kraus Operators for BB84} \label{apd:bb84ops}
We compute the Kraus operators from \ref{apd:kraus}. In general Alice's POVM elements are
\begin{align}
    \Gamma^A_{Z,0} &=\Gamma^A_{1} =\sum_{n=0}^{K+1}
    \ketbra{n}\otimes\ketbra{1}\notag
    \\
    \Gamma^A_{Z,1} &=\Gamma^A_{2} =\sum_{n=0}^{K+1}
    \ketbra{n}\otimes\ketbra{2}\notag
    \\
    \Gamma^A_{X,0} &=\Gamma^A_{3} =\sum_{n=0}^{K+1}
    \ketbra{n}\otimes\ketbra{3}\notag
    \\
    \Gamma^A_{X,1} &=\Gamma^A_{4} =\sum_{n=0}^{K+1}
    \ketbra{n}\otimes\ketbra{4}\notag
    \\
\end{align}
Note that the PNS attack allows Eve to gain full information of signals with multi-photon components. Therefore in BB84, we set the photon number cutoff to be $K = 1$. Bob's POVM elements, as they lie in the $n\leq 1$ photon space, are 
\begin{align}
    \Gamma^B_{Z,0} &=\Gamma^B_{H}= p_Z\ketbra{0}{0} = p_Z\begin{pmatrix} 1 & 0 & 0 \\ 0 & 0 & 0 \\ 0 & 0 & 0\end{pmatrix},\notag \\
    \Gamma^B_{Z,1} &=\Gamma^B_{V}=p_Z\ketbra{1}{1} = p_Z\begin{pmatrix} 0 & 0 & 0 \\ 0 & 1 & 0 \\ 0 & 0 & 0\end{pmatrix},\notag \\
    \Gamma^B_{X,0} &=\Gamma^B_{+}=p_X\ketbra{+}{+} = \frac{1-p_Z}{2}\begin{pmatrix} 1 & 1 & 0 \\ 1 & 1 & 0 \\ 0 & 0 & 0\end{pmatrix},\notag \\
    \Gamma^B_{X,1} &=\Gamma^B_{-}= p_X\ketbra{-}{-} = \frac{1-p_Z}{2}\begin{pmatrix} 1 & -1 & 0 \\ -1 & 1 & 0 \\ 0 & 0 & 0\end{pmatrix},\notag\\
    \Gamma^B_{\perp} &=\ketbra{\perp}{\perp} = \begin{pmatrix} 0 & 0 & 0 \\ 0 & 0 & 0 \\ 0 & 0 & 1\end{pmatrix}.
\end{align}
Note we discard basis mismatched announcements and vacuum signals. We apply isometries to eliminate redundant registers. The resulting Kraus operators are
\begin{align}
    K_Z:=K_{\alpha=Z,\beta=Z} =& \left(\ket{0}_R\otimes\sqrt{\Gamma^A_{Z,0}}+\ket{1}_R\otimes\sqrt{\Gamma^A_{Z,1}}\right)\notag\\
    &\otimes \sqrt{p_Z} \left(\mathds{1}_{2\times 2}\oplus 0\right)_B\otimes\ket{0}_C,
\end{align}
and
\begin{align}
    K_X:=K_{\alpha=X,\beta=X} =& \left(\ket{0}_R\otimes\sqrt{\Gamma^A_{X,0}}+\ket{1}_R\otimes\sqrt{\Gamma^A_{X,1}}\right)\notag\\
    &\otimes \sqrt{p_X} \left(\mathds{1}_{2\times 2}\oplus 0\right)_B\otimes\ket{1}_C,
\end{align}
where $C$ is a two dimensional register, resulting in a $\mathcal{G}$ map
\begin{align}
    \mathcal{G}(\rho) = K_Z\rho K_Z^\dagger + K_X\rho K_X^\dagger .
\end{align}
\vspace{-1cm}

\section{Kraus Operators for NPAB BB84}
Alice and Bob perform the same measurements and so the resulting POVM elements are the same as in Appendix \ref{apd:bb84ops}. However, only a single announcement is made announcing the   signal detected. The Kraus operator is thus
\begin{align}
    K_N :=& K_{\beta=\text{detect}} \notag
    \\
    =& \bigg(\ket{0}_R\otimes\ket{0}_X\otimes\sqrt{\Gamma^A_{Z,0}}+\ket{1}_R\otimes\ket{0}_X\otimes\sqrt{\Gamma^A_{Z,1}}\notag
    \\
    &+\ket{0}_R\otimes\ket{1}_X\otimes\sqrt{\Gamma^A_{X,0}}+\ket{1}_R\otimes\ket{1}_X\otimes\sqrt{\Gamma^A_{X,1}}\bigg)\notag
    \\
    &\otimes \left(\mathds{1}_{2\times 2}\oplus 0\right)_B,
\end{align}
where the register $X$ now encodes Alice's basis choice, and the register $C$ is one dimensional and thus trivial. The resulting $\mathcal{G}$ map is
\begin{align}
    \mathcal{G}(\rho) = K_N\rho K_N^\dagger.
\end{align}

\section{Kraus Operators for SARG04}
Alice and Bob again perform the same measurements and so the resulting POVM elements are the same as in Appendix \ref{apd:bb84ops}. The announcement structure of SARG04 yields multiple Kraus operators for the $\mathcal{G}$ map. As only the $\beta = 1$ announcement survives sifting, the Kraus operators are
\begin{align}
    K_1:=&K_{\alpha = 1}=\bigg(\ket{0}_R\otimes\sqrt{\frac{1}{2}\Gamma^A_{Z,0}}\otimes\sqrt{1-p_Z}\ketbra{-}{-}_B \notag\\
    &+\ket{1}_R\otimes\sqrt{\frac{1}{2}\Gamma^A_{X,0}}\otimes\sqrt{p_Z}\ketbra{1}{1}_B
    \bigg)\otimes\ket{1}_C,\notag
    \\
    K_2:=&K_{\alpha = 2}=\bigg(\ket{0}_R\otimes\sqrt{\frac{1}{2}\Gamma^A_{Z,1}}\otimes\sqrt{1-p_Z}\ketbra{-}{-}_B \notag\\
    &+\ket{1}_R\otimes\sqrt{\frac{1}{2}\Gamma^A_{X,0}}\otimes\sqrt{p_Z}\ketbra{0}{0}_B
    \bigg)\otimes\ket{2}_C,\notag
        \\
    K_3:=&K_{\alpha = 3}=\bigg(\ket{0}_R\otimes\sqrt{\frac{1}{2}\Gamma^A_{Z,1}}\otimes\sqrt{1-p_Z}\ketbra{+}{+}_B \notag\\
    &+\ket{1}_R\otimes\sqrt{\frac{1}{2}\Gamma^A_{X,1}}\otimes\sqrt{p_Z}\ketbra{0}{0}_B
    \bigg)\otimes\ket{3}_C,\notag
        \\
    K_4:=&K_{\alpha = 4}=\bigg(\ket{0}_R\otimes\sqrt{\frac{1}{2}\Gamma^A_{Z,0}}\otimes\sqrt{1-p_Z}\ketbra{+}{+}_B \notag\\
    &+\ket{1}_R\otimes\sqrt{\frac{1}{2}\Gamma^A_{X,1}}\otimes\sqrt{p_Z}\ketbra{1}{1}_B
    \bigg)\otimes\ket{4}_C.
\end{align}
The resulting $\mathcal{G}$ map is
\begin{align}
    \mathcal{G}(\rho) = \sum_{i=1}^4K_i\rho K_i^\dagger.
\end{align}
Note that each of Alice's POVMs has a additional $\frac{1}{2}$ to indicate the fact that given Alice is sending a particular polarisation, Alice randomly chooses one of the two different equal probable announcements and associates it with this signal. 
\end{document}